\documentclass{elsart} 

\usepackage{epsfig}
\usepackage{latexsym,amssymb,amsmath}

\journal{Phys.Lett.B}

\newcommand{\eq}[1]{\begin{equation} #1 \end{equation}}
\newcommand{\GeV}{\mbox{$\,{\rm GeV}$}}

\begin{document}

\begin{frontmatter}

\title{Multiplicity Fluctuations
      in Proton-Proton and Nucleus-Nucleus
      Collisions}

\author[bitp,hrs]{V.P.~Konchakovski},
\author[bitp,fias]{M.I.~Gorenstein}, and
\author[fias]{E.L.~Bratkovskaya}

\address[bitp]{Bogolyubov Institute for Theoretical Physics, Kiev, Ukraine}
\address[hrs]{Helmholtz Research School, University of Frankfurt, Frankfurt, Germany}
\address[fias]{Frankfurt Institute for Advanced Studies, Frankfurt, Germany}

\begin{abstract}
We study the scaled variances of multiplicity fluctuations in
nucleus-nucleus collisions at SPS and RHIC energies within the HSD
transport model. The HSD results are compared with
proton-proton data and with predictions of the hadron-resonance
gas statistical model. We find that the HSD scaled variances
$\omega_i$ for negatively, positively, and all charged hadrons in
central nucleus-nucleus collisions remain close to the $\omega_i$
values in proton-proton collisions and increase with
collision energy as the corresponding multiplicities per
participating nucleon. The statistical model  predicts  very
different behavior of $\omega_i$. However, a comparison with
preliminary NA49 data for the most central Pb+Pb collisions at SPS
energies does not permit to distinguish  the HSD and statistical
model results. New measurements of the multiplicity fluctuations
in nucleus-nucleus collisions in a wide energy region with large
acceptance are needed to allow for a proper determination of the
underlying dynamics.
\end{abstract}


\end{frontmatter}


The event-by-event fluctuations in high energy nucleus-nucleus
(A+A) collisions (see e.g., the reviews \cite{rev1,rev2}) are
expected to be closely related to the transition between
different phases of the QCD matter. Measuring the fluctuations one
might observe anomalies of the onset of deconfinement \cite{ood}
and dynamical instabilities when the expanding system goes through
the 1-st order transition line between the quark-gluon plasma and
the hadron gas \cite{fluc2}. Furthermore, the QCD critical point
may be signaled by a characteristic pattern in fluctuations
\cite{fluc3}. In the present paper we calculate the multiplicity
fluctuations in central A+A collisions at SPS and RHIC energies
within the microscopic Hadron-String-Dynamics~(HSD) transport
model \cite{HSD} which gives  rather reliable description for the
inclusive spectra of charged hadrons in A+A collisions from SIS to
RHIC energies \cite{Weber}.

\begin{figure}[t!]
\centerline{\epsfig{file=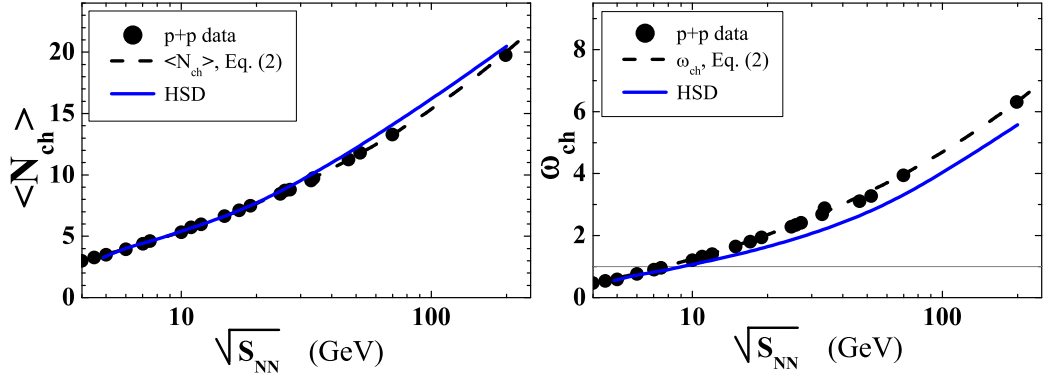,width=\textwidth}}
\caption{ The multiplicity {\it (left)} and scaled variance {\it
(right)} of all charged hadrons in p+p inelastic collisions as
functions of collision energy. The dashed lines correspond to the
parametrization (\ref{Nchpp}) from Ref.~\cite{rev1}. The solid
lines are the HSD results.} \label{flucP}
\end{figure}

For a quantitative measure of the particle number fluctuations it
is convenient to use the scaled variances,
\eq{
 \omega_i~\equiv~\frac{\langle N_i^2\rangle~-~\langle N_i\rangle^2}{\langle N_i\rangle}~,
 \label{omega_def}
}
where $\langle \cdots\rangle$ denotes event-by-event averaging
and the index $i$ means ``-'', ``+'', and ``ch'', i.e negative,
positive, and all charged final state hadrons. The compilation of
proton-proton (p+p) data for $\langle N_{ch}\rangle$ and
$\omega_{ch}$ are taken from Ref.~\cite{rev1} and presented in
Fig.~\ref{flucP}. The energy dependence can be parameterized by
the functions \cite{rev1}:
 \eq{\label{Nchpp}
\langle N_{ch}\rangle~\cong~
-4.2~+~4.69~\left(\frac{\sqrt{s_{NN}}}{\GeV}\right)^{0.31}~,~~~~
%
%
%
%
%
 \omega_{ch}~\cong~ 0.35~\frac{(\langle
N_{ch}\rangle~-~1)^2}{\langle N_{ch}\rangle}~,
}
where $\sqrt{s_{NN}}$ is the center of mass energy.

The HSD model description of the p+p data (for p+p reaction this
is almost equivalent to the Lund-String model \cite{Lund}) is shown in
Fig.~\ref{flucP} by the solid lines. It gives a good reproduction of
the p+p data for $\langle N_{ch}\rangle$, but slightly
underestimates $\omega_{ch}$ at high collision energies. For
negative and positive charged hadrons the average
multiplicities and scaled variances  in p+p collisions can be
presented in terms of the corresponding quantities for all charged
particles,
\eq{ \label{posneg}
\langle N_{\pm}\rangle~=~\frac{1}{2}~ \left(\langle
N_{ch}\rangle~\pm~ 2\right)~,~~~~ \omega_{\pm}~=~\frac{1}{2}~
\omega_{ch} ~\frac{\langle N_{ch}\rangle}{\langle
N_{ch}\rangle~\pm~ 2}~.
}

In general, one can define two groups of hadron observables. The first
group includes observables which are rather similar in  A+A and p+p
collisions, thus, they can be reasonably described within the wounded
nucleon model (WNM) \cite{WNM} which treats the final state in A+A
collisions as the result of independent nucleon-nucleon (N+N)
collisions.  The second group consists of A+A observables which are
very different from those in p+p collisions (e.g. collective flow,
multi-strange baryon production).
The question arises: Are the multiplicity fluctuations in A+A
collisions close to those in p+p reactions, or are they very different?

\begin{figure}[t!]
\centerline{\epsfig{file=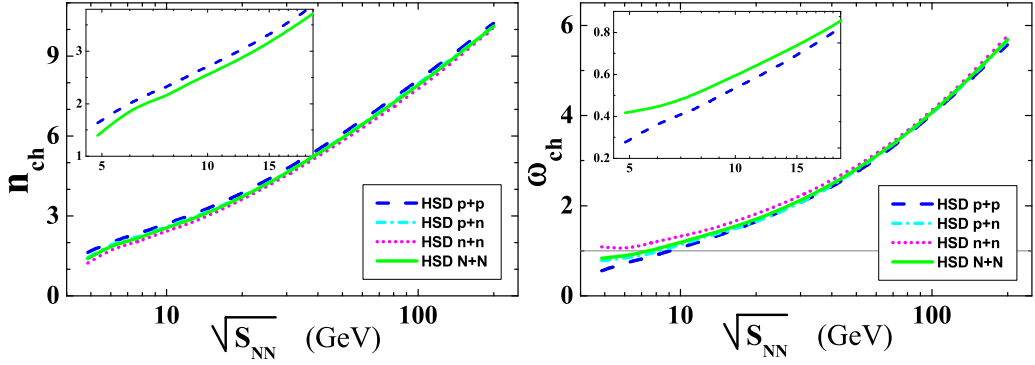,width=\textwidth}}
\caption{ The HSD results for $n_{ch}$ ({\it left}) and
$\omega_{ch}$ ({\it right}) in p+p (dashed line), p+n
(dashed-dotted line), n+n (dotted line), and N+N (solid line)
collisions.} \label{NN}
\end{figure}

To answer this question let us first consider the model
predictions. To compare central collisions of heavy nuclei and N+N
collisions within the HSD model we construct the multiplicities
and scaled variances of N+N reactions using the HSD results for
p+p, p+n and n+n collisions:
\eq{ \label{mult-NN}
\langle N_i^{NN}\rangle~=~\alpha_{pp}~\langle
N_i^{pp}\rangle~+~\alpha_{pn}~\langle
N_i^{pn}\rangle~+~\alpha_{nn}~\langle N_i^{nn}\rangle~, }
%
\eq{ \label{omega-NN}
\omega_i^{NN}~=~\frac{1}{\langle
N_i^{NN}\rangle}~\left[\alpha_{pp}~\omega_i^{pp}~\langle
N_i^{pp}\rangle~+~ \alpha_{pn}~\omega_i^{pn}~\langle N_i^{pn}\rangle~
+~\alpha_{nn}~\omega_i^{nn}~\langle N_i^{nn}\rangle\right]~,
}
where $\alpha_{pp}=Z^2/A^2\cong 0.16,~\alpha_{pn}=2Z(A-Z)/A^2\cong
0.48,~\alpha_{nn}=(A-Z)^2/A^2\cong 0.36$ are the probabilities of
proton-proton, proton-neutron, and neutron-neutron collisions in
Pb+Pb (A=208, Z=82) or Au+Au (A=197, Z=79) reactions. The results
for $p+p, p+n$ and $n+n$ collisions -- presented in Fig.~\ref{NN} --
are very close to each other.
A small difference between p+p and N+N collisions ($<5$\%) is only seen
at SPS energies (shown separately in the upper left corners) and
gradually disappears at  RHIC energies.

\begin{figure}[t!]
\centerline{\epsfig{file=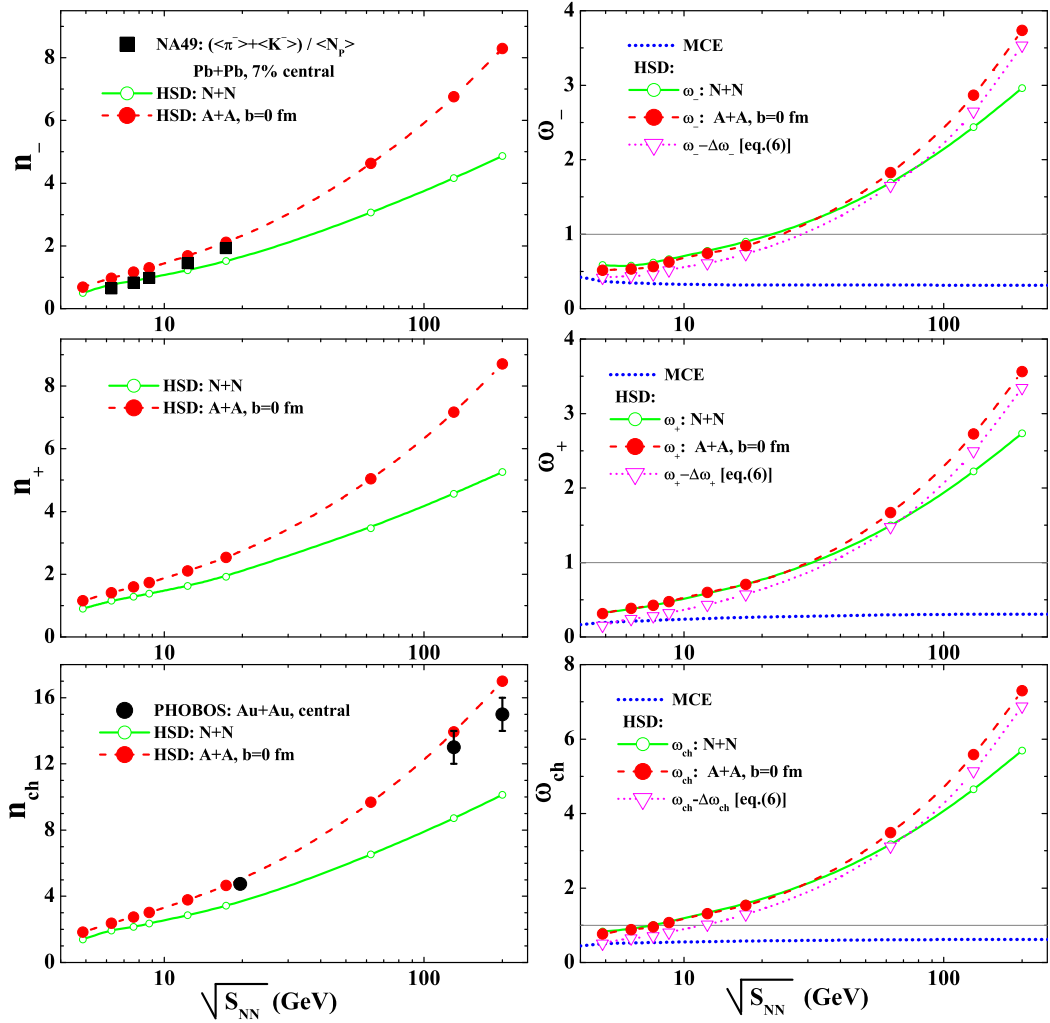,width=\textwidth}}
\caption{The multiplicities per participant, $n_i$ ({\it left}),
and scaled variances, $\omega_i$ ({\it right}). The solid lines
are the HSD results for N+N collisions according to
Eq.~(\ref{mult-NN}). The full circles are the HSD results for
central A+A collisions for zero impact parameter, $b=0$.  The full
squares for $n_-$ are the NA49 data \cite{data} for $(\langle
\pi^-\rangle +\langle K^-\rangle)/\langle N_P\rangle$  in the
samples of 7\% most central Pb+Pb collisions.
The full black dots for $n_{ch}$ are the PHOBOS data \cite{PHOBOSnch}.
The HSD results for
$\omega_i$ after the subtraction of the contributions $\Delta
\omega_i$ (\ref{delta-omega})  are shown by open triangles. The
dotted lines are the MCE HG model results for
$\omega_i$~\cite{res_MCE}. The HG parameters correspond to the
chemical freeze-out conditions  found from fitting the hadron
yields.} \label{fluc-A}
\end{figure}

In Fig.~\ref{fluc-A} the HSD model results are shown for the
multiplicities per participating nucleons, $n_i=\langle
N_i\rangle/\langle N_P\rangle$, and for the scaled variances,
$\omega_i$, in central collisions (zero impact parameter, $b=0$)
of Pb+Pb at $E_{lab}=10$, 20, 30, 40, 80, 158~AGeV and  Au+Au at
$\sqrt{s_{NN}}=62$, 130, 200~GeV. From Fig.~\ref{fluc-A} one
concludes that the HSD results for the scaled variances in central
A+A collisions are close to those in N+N collisions. For the SPS
energy region all scaled variances, $\omega_{\pm}$ and
$\omega_{ch}$, in central  A+A collisions are slightly below the
N+N results. The reversed situation is observed for RHIC energies.
Thus, the HSD results for multiplicity fluctuations are rather
similar to those of the WNM. This is in agreement with our
previous finding \cite{voka1}. For the samples with a fixed number
of nucleon participants, $N_P^{proj}=N_P^{targ}=const$, in Pb+Pb
collisions at 158~AGeV, HSD shows fluctuations of the final
hadrons close to those in N+N collisions at the same energy. This
happens to be also valid for most central collisions ($b=0$)
considered in the present study. To estimate the influence of
participant number fluctuations we calculate the scaled variance,
$\omega_P=Var(N_P)/\langle N_P\rangle$, in the HSD simulations.
The participant number fluctuations are found to be rather small
for collisions with $b=0$. For example, in Pb+Pb collisions with
$b=0$ at 158~AGeV the mean number of participants is
$\langle N_P\rangle \cong 392$, and the scaled variance is
$\omega_P\cong 0.055$ . The additional fluctuations, $\Delta
\omega_i$, of $i$th hadrons due to participant number fluctuations
can be estimated as \cite{voka1},
 \eq{ \label{delta-omega}
\Delta \omega_i ~=~ n_i~\omega_P~.
}
The HSD results for $\omega_i$ after subtraction of the
contributions $\Delta \omega_i$ (\ref{delta-omega})  are shown in
Fig.~\ref{fluc-A} by open triangles. The contributions to
$\omega_i$ due to participant number fluctuations estimated by
Eq.~(\ref{delta-omega}) are small, and they do not explain the
(positive) difference, $\omega_i($AA$)-\omega_i($NN$)$ seen in
Fig.~\ref{fluc-A} at $\sqrt{s_{NN}}=200$~GeV.

In the statistical model the scaled variances
$\omega_i = 1$ for the ideal Boltzmann gas in the grand canonical ensemble
(GCE). The deviations of $\omega_i$ from unity in the
hadron-resonance gas (HG) model stem from Bose and Fermi
statistics, resonance decays, and exactly enforced conservations
laws within the canonical ensemble (CE) or micro-canonical
ensemble (MCE) \cite{res_MCE,CE}. Note that the statistical model
gives no predictions for the energy dependence of hadron
multiplicities. All yields are proportional to the system volume
$V$ which is a free model parameter fitted to the multiplicity
data at each collision energy. However, the statistical model does
predict the scaled variances as $\omega_i$ to become independent of
the system volume for large systems. In Fig.~\ref{fluc-A} the
scaled variances $\omega_i$ calculated within the MCE HG model
along the chemical freeze-out line (see Ref.~\cite{res_MCE} for
details) are presented by the dotted lines:  $\omega_i$ reach
their asymptotic values at RHIC energies,
$\omega_{\pm}$(MCE)$\cong 0.3$ and $\omega_{ch}$(MCE)$\cong 0.6$.
The corresponding results in the GCE and CE are the following:
$\omega_{\pm}$(GCE)$\cong 1.2$ and $\omega_{ch}$(GCE)$\cong 1.6$,
$\omega_{\pm}$(CE)$\cong 0.8$ and $\omega_{ch}$(CE)$\cong 1.6$.
The HSD results for $\omega_i$ in central A+A collisions are very
different. They remain close to the corresponding values in p+p
collisions and, thus, increase with collision energy as
$\omega_i\propto n_i$. One observes no indication for
`thermalization' of fluctuations in the HSD results. This is
especially seen for RHIC energies:
$\omega_{i}$(HSD)$/\omega_i$(MCE)$\ge 10$ at
$\sqrt{s_{NN}}=200$~GeV.

Recently first measurements of fluctuations of particle
multiplicity \cite{fluc-mult1,fluc-mult1a,fluc-mult2} and
transverse momenta \cite{fluc-pT} in A+A collisions have been
performed. The scaled variance for negatively, positively, and all
charged hadrons was measured as a function of centrality at SPS
\cite{fluc-mult1,fluc-mult1a} and RHIC \cite{fluc-mult2} energies.
It has been argued in Ref.~\cite{voka1} that the final
multiplicity fluctuations seen in the NA49 data \cite{fluc-mult1}
at the SPS are a straightforward consequence of the fluctuations
in the number of nucleon participants (see also the discussion in
Ref.~\cite{MGMG}). It seems that the same conclusion \cite{voka2}
is valid for the RHIC data \cite{fluc-mult2}.   In the language of
statistical models, the fluctuations of the number of nucleon
participants correspond to volume fluctuations, hence, they
translate directly to the final multiplicity fluctuations. To
avoid these `trivial' fluctuations, one has to select a sample of
very central, $\leq 1 \%$, collisions. Such a rigid centrality
selection has been recently done for the NA49 data \cite{lungwitz}
by fixing the number of projectile participants, $N_P^{proj}\cong
A$.

\begin{figure}[t!]
\centerline{\epsfig{file=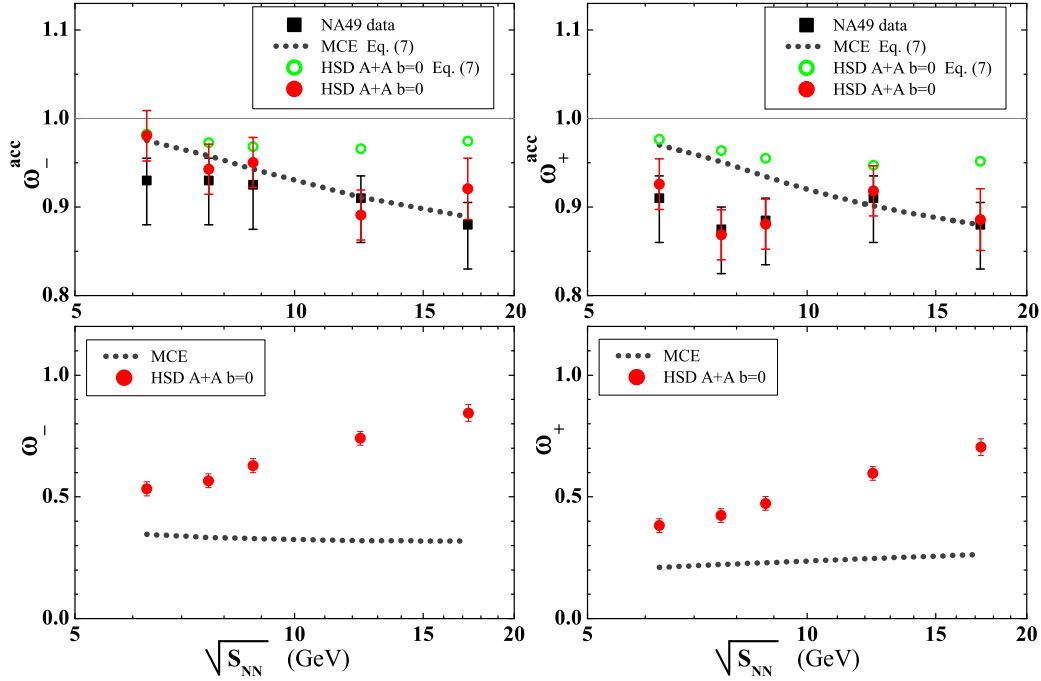,width=\textwidth}}
\caption{{\it Upper panel.} The scaled variances
$\omega^{acc}_{\pm}$ for  central Pb+Pb collisions. The squares
are the NA49 data for  1\% most central collisions
\cite{lungwitz}.
The dotted lines show the MCE HG model results
calculated from full 4$\pi$ scaled variances using
Eq.~(\ref{acc}). The full circles present the HSD results
in Pb+Pb collisions for $b=0$
with the NA49 experimental acceptance conditions, while the open
circles are obtained from the 4$\pi$ HSD scaled variances using
Eq.~(\ref{acc}). {\it Lower panel.} The MCE HG (dotted line) and
HSD (full circles) results for the 4$\pi$ scaled variances
$\omega_{\pm}$ (the same as in Fig.~\ref{fluc-A}) are shown for
SPS energies. } \label{NA49}
\end{figure}

The HG model was compared with the NA49 data~\cite{lungwitz} for
the sample of 1\% most central collisions at the SPS energies,
$20\div 158$~AGeV in Ref.~\cite{res_MCE}. It was found  that the
MCE results for $\omega_{\pm}$ are very close to the data, they
are shown by the dashed lines in Fig.~\ref{NA49}. The
NA49 acceptance probabilities for positively and negatively
charged hadrons are approximately equal, and their numerical
values are: $q=0.038$, 0.063, 0.085, 0.131, 0.163, at the SPS
energies of 20, 30, 40, 80, 158~AGeV, respectively. In the
statistical model the scaled variances $\omega_{\pm}^{acc}$ for
the accepted particles are calculated from $\omega_{\pm}$ in the
full space according to the acceptance scaling formulae (see
Ref.~\cite{res_MCE} for details):
\eq{\label{acc}
\omega_{\pm}^{acc}~=~1~-~q~+~q~\omega_{\pm}~.
}
Note that the energy dependence of $\omega^{acc}_{\pm}$ seen in
Fig.~\ref{NA49} is strongly influenced by an increase with energy
of the acceptance parameter $q$: only about 4\% of the hadrons are
detected at 20~AGeV and 16\% at 158~AGeV.

The comparison of the HSD results for central Pb+Pb collisions
(zero impact parameter, $b=0$) with the preliminary NA49 data of
1\% most central collisions, selected by the number of projectile
spectators, is presented in Fig.~\ref{NA49}. It demonstrates a
good agreement of the HSD results with the preliminary NA49 data.
There are also no essential differences between the MCE HG model
and the HSD transport model results. Several comments are needed
at this point:
The HSD results for  $\omega_{\pm}^{acc}$ calculated within the NA49
acceptance  (full circles in Fig.~\ref{NA49}) lead to smaller
values of $\omega_{\pm}^{acc}$ than those obtained with the acceptance
scaling formulae (\ref{acc}) (open circles in Fig.~\ref{NA49}), i.e. to the
violation of the acceptance scaling formulae (\ref{acc}).  This difference
may lead to a 10\% effect in $\omega_{\pm}^{acc}$ for the NA49 acceptance
conditions. Thus, the MCE results for $\omega_{\pm}^{acc}$ may also be about
10\% smaller than those obtained from Eq.~(\ref{acc})  shown in the upper
panel of Fig.~\ref{NA49}. 

The lower panel of Fig.~\ref{NA49} demonstrates that the MCE and HSD 
results for $\omega_{\pm}$ at the lowest  SPS energy 20~AGeV are 
accidentally rather close to each other. They both are also close to 
$\omega_{\pm}$ in p+p collisions (cf.  Figs.~\ref{NN} and 
\ref{fluc-A}). The HSD scaled variances $\omega_{i}$ increase with 
collision energy. In contrast, the MCE $\omega_i$ values remain 
approximately constant. The ratio of the HSD to MCE values of 
$\omega_{\pm}$ reaches about the factor of 2 at the highest SPS energy 
158~AGeV. It becomes a factor of 10 at the top RHIC energy 
$\sqrt{s_{NN}}=200$~GeV. However, the rigid centrality selection is 
absent for the available RHIC fluctuation data. Due to this reason the 
participant number fluctuations give a dominant contribution to 
$\omega_i$. On the other hand, for the SPS data the small values of the 
acceptance, $q=0.04\div 0.16$, and 10\% possible ambiguities coming 
from Eq.~(\ref{acc}) almost mask the difference between the HSD and MCE 
results (Fig.~\ref{NA49}, upper panel).

In summary, we have used the HSD transport model to study the particle
number fluctuations in central A+A collisions at SPS and RHIC
energies.   HSD predicts that the scaled variances $\omega_i$
in central A+A collisions remain close to the corresponding values
in p+p collisions and  increase with collision energy as the
multiplicity per participating nucleon, i.e. $\omega_i\propto n_i$. The
scaled variances $\omega_i$ calculated within the statistical HG model
along the chemical freeze-out line show a rather different behavior:
$\omega_i$ approach finite values at high collision energy. At the top
RHIC energy, $\sqrt{s_{NN}}=200$~GeV, the HSD values of
$\omega_{i}$(HSD) is already about 10 times larger than the
corresponding MCE HG values of $\omega_i$(MCE).

The HSD and HG scaled variances $\omega_i$ show a different
energy dependence such that the deviation grows with increasing energy.
However, a comparison with preliminary NA49 data of very central, $\leq
1 \%$, Pb+Pb collisions at the SPS energy range does not distinguish
between the HSD and MCE HG results due to two
reasons: First, the MCE HG and HSD results for $\omega_{i}$ at SPS
energies are not too much different from each other and from
$\omega_{i}$ in p+p collisions. Second, small experimental coverage of
the acceptance, $q=0.04\div 0.16$, and 10\% possible ambiguities coming
from the acceptance scaling relation (\ref{acc}) make the difference
between the HSD and MCE HG results almost invisible.

Thus, our study shows a clear way to distinguish the different models
by measuring the multiplicity fluctuations.   However, the later are
very sensitive to the experimental acceptance, such that  new
measurements of $\omega_i$ for very central A+A collisions with large
acceptance at both SPS \cite{NA49future} and RHIC energies are required
to allow for a proper determination of the underlying dynamics.

{\bf Acknowledgments. }\; We would like to thank V.V.~Begun,
W.~Cassing, M.~Ga\'zdzicki, W.~Greiner, M.~Hauer, B.~Lungwitz
and H. St\"ocker
for useful  discussions. One of the author (M.I.G.) is thankful to
the Humboldt Foundation for financial support.


\end{document}